# ADNAC: Audio Denoiser using Neural Audio Codec

Daniel Jimon, Mircea Vaida, Adriana Stan
Technical University of Cluj-Napoca, Romania
jimon.mi.lucian@student.utcluj.ro, {mircea.vaida, adriana.stan}@com.utcluj.ro

*Abstract*—Audio denoising is critical in signal processing, enhancing intelligibility and fidelity for applications like restoring musical recordings. This paper presents a proof-of-concept for adapting a state-of-the-art neural audio codec, the Descript Audio Codec (DAC), for music denoising. This work overcomes the limitations of traditional architectures like U-Nets by training the model on a large-scale, custom-synthesized dataset built from diverse sources. Training is guided by a multi-objective loss function that combines time-domain, spectral, and signal-level fidelity metrics. Ultimately, this paper aims to present a PoC for high-fidelity, generative audio restoration.

## I. INTRODUCTION

Noise reduction is a fundamental part of audio signal processing, substantially improving signal quality and intelligibility across domains like speech processing [1-3], music production and restoration [1], and bioacoustics analysis [2]. Even though traditional techniques, such as spectral subtraction and Wiener filtering, provide a foundational basis, they often fail with complex and non-stationary noise patterns.

Recent advancements in deep learning, particularly with convolutional neural networks (CNNs) in encoder-decoder structures like the U-Net, have shown significant promise [3, 4]. Previous works demonstrated that a simple U-Net architecture can effectively filter white and urban noise from musical signals but fails to address more complex degradations like reverberation and noise cancellation artifacts. These limitations highlight the need for more powerful models capable of understanding and regenerating audio with higher fidelity.

This paper investigates a novel approach which goes beyond conventional discriminative models by adapting a generative neural audio codec for denoising. The goal was to leverage the Descript Audio Codec (DAC) [5], a model designed for high-fidelity audio compression and retrain it on a denoising objective. The core hypothesis is that the rich representations and generative capabilities of a neural codec can produce better denoising results, minimizing artifacts and preserving the natural aspects of the original music, more efficiently than a standard U-Net. This is especially true when a multi-objective loss function is used, and the training is performed on a comprehensive, purpose-built dataset.

## II. RELATED WORK

The field of audio denoising has been largely driven by progress in speech enhancement, like how deep learning benefited from image processing advancements. Established deep learning methods include systems based on CNN-based models [6], Long Short-Term Memory (LSTM) networks [7], autoencoder configurations [8] etc. More recently, state-of-the-art approaches have been tried with transformer-based architectures like generative adversarial networks (GANs) like CMGAN [9] to enhance perceptual quality or DPT-FSNet for improved feature extraction [10].

In addition, the current state of the art in many high-fidelity generative modelling tasks is dominated by Denoising Diffusion Probabilistic Models (DDPMs) [11,12]. Diffusion models work by learning to reverse a gradual noising process. [11].

While this iterative refinement enables diffusion models to generate exceptionally high-quality audio [12], the process is computationally intensive and slow, often requiring hundreds or even thousands of sequential passes through a large neural network to generate a single piece of audio.

A significant research gap exists in applying these techniques specifically to music. While GANs have improved perceptual quality by using discriminators, the presented work explores a different approach by using neural audio codecs. Neural audio codecs such as SoundStream [13], EnCodec [14] and Descript Audio Codec (DAC) [5] represent the newest focus in audio compression. These architectures learn to transform the audio into a compact, quantized form, which then can be used for a high-fidelity waveform reconstruction. The potential for audio enhancement tasks has started to be recognized. This paper is based on this idea by repurposing a pre-trained DAC model. This is performed by fine-tuning its generative decoder using a denoising objective guided by a mix of reconstruction and perceptual losses.

## III. DATASET AND DEGRADATION

A robust and diverse dataset is of primary importance for a machine learning training purpose, thus the presented denoising model needed a proper amount of useful data. A comprehensive dataset was synthesized specifically for music denoising, using automated scripts to ensure reproducibility. All audio files were processed into 2-seconds mono chunks at a 44.1kHz sample rate.

### A. Clean Music Sources

A base dataset of high-quality, varied audio was aggregated from multiple standard dataset:

*1) IRMAS Dataset [15]*: A renowned musical dataset containing 6705 instrumental clips used in several papers, offering a solid foundation because of the variety of instruments used.

*2) M&N Dataset (Music) [16]*: Contains 3.43 hours of clean music recordings, including sounds from instruments like piano, drums, harp etc.

*3) MUSAN Dataset (Music) [17]*: Includes longer musical pieces, providing contextual variety.

### B. Degradation Sources

To simulate a wide range of real-world audio problems, a diverse set of noise and reverberation sources was compiled:

*1) Additive Noise:* Programatically created white noise, manually chosen 44.1kHz urban noise snippets [18] and a

rich selection of environmental noises from the M&N [16] and FSD50K [19] datasets.

*2) Reverberation:* Simulated reverb with the Pedalboard Python module [20], and convolution-resulted reverberation effects, by using the BUTReverbDB dataset's RIR examples [21].

*3) Noise Cancellation Artifacts:* A custom algorithm was used to simulate artifacts from low-quality consumer devices by applying random partial attenuations and frequency suppression to the clean signals.

Additive noise (white, urban, environmental) represents the most common form of audio degradation and serves as a baseline challenge for any denoising system. Reverberation was included because it represents a convolutional, rather than additive, distortion that traditional filters and simple neural networks often fail to mitigate effectively, smearing temporal details and reducing clarity. Finally, the inclusion of simulated noise cancellation (NC) artifacts addresses a modern challenge, where aggressive but low-quality processing on consumer devices introduces characteristic non-linear distortions that are difficult to model and reverse.

*C. Dataset Generation Pipeline*

An automated script was developed to create the final training, validation, and test sets. For each clean audio chunk, the script creates multiple degraded versions by randomly selecting one or two degradations, applying them sequentially (e.g., adding reverb, then mixing with noise), and mixing noise at a random Signal-to-Noise Ratio (SNR) between 0dB and 15dB. This process generated a large-scale paired dataset, with a metadata file logging every clean-noisy pair, the degradations applied, and the SNR level, ensuring full experimental reproducibility.

IV. MODEL ARCHITECTURE AND ADAPTATION

The proposed model leverages a pre-trained neural audio codec as a foundation and introduces a denoising network that operates entirely within the codec's latent space. The overall architecture first encodes the noisy audio into a compact latent representation using the DAC encoder. This representation is then fed into a denoiser U-Net, which is trained to predict clean latent representation. Finally, the denoised latent representation is passed through the frozen DAC decoder to synthesize the clean audio waveform. The codes and scripts used are available at *https://github.com/jimonld2000/ADNAC*.

*A. Descript Audio Codec (DAC)*

DAC is a high-fidelity neural audio codec based on a RVQGAN (Residual Vector Quantized Generative Adversarial Network) [5]. The Descript Audio Codec (DAC) has been positioned as a state-of-the-art audio tokenizer, reportedly offering improvements over earlier neural codecs such as SoundStream [13] and EnCodec [14]. Developed by Descript, DAC is designed for high-fidelity audio representation and boasts a significant compression factor. It supports a range of common audio sampling rates, including 44.1 kHz, 24 kHz, and 16 kHz, and can handle both monophonic and stereophonic audio signals [22].

A key advantage of DAC, particularly for rapid prototyping and integration, is its ease of use. The Hugging Face model *hance-ai/descript-audio-codec-44khz* [23] provides a straightforward Python API, allowing developers to encode audio files into either discrete embeddings (denoted as zq) or token sequences (s), and subsequently decode these representations back into audio waveforms, often with just a single line of code for each operation [23]. This simplicity lowers the barrier to entry for leveraging its advanced tokenization capabilities.

The primary components of the DAC architecture are an Encoder that maps a raw audio waveform to a lower-dimensional continuous latent representation; a Quantizer that discretizes this representation into a sequence of codes; and a Decoder that synthesizes a high-fidelity output waveform from these codes. A key architectural detail is its use of 9 layers of Residual Vector Quantization (RVQ) with 10-bit codebooks, which improves codebook usage and reconstruction quality.

*B. Latent Denoiser U-Net*

The core of the proposed model is a U-Net that operates on the 1D latent sequences produced by the DAC encoder. The main building block is a residual block containing Group Normalization, SiLU activations, and 1D convolutions. Residual connections are essential for training deeper networks. The model has an encoder path that downsamples the latent sequence and a decoder path that upsamples it, with skip connections linking corresponding levels. This allows the model to process the latent representation at multiple temporal resolutions while preserving sequence length.

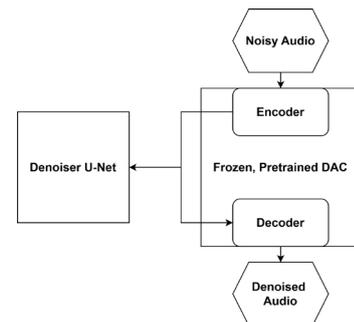

Fig. 1. The architecture of the proposed latent-domain denoising model.

The model's configuration can be seen in Fig. 1. It first encodes the noisy audio into a latent representation using the DAC encoder. This latent representation is then normalized and fed into our Latent Denoiser U-Net, which is trained to predict the clean latent representation. Finally, the denoised latent representation is passed through the DAC decoder to synthesize the clean audio waveform.

*C. Architectural and Training Details*

The Latent Denoiser U-Net was designed to be both powerful and efficient. The encoder path consists of five downsampling blocks, each containing two residual convolutional blocks followed by a strided convolution with a kernel size of 4 to halve the temporal resolution [22]. The decoder path mirrors this structure with five upsampling blocks using transposed convolutions. Skip connections concatenate the feature maps from corresponding encoder and decoder levels, allowing the network to combine high-level semantic information with low-level temporal details [24]. The number of channels begins at 64 and doubles with each downsampling step, capping at 512 in the bottleneck layer.

The decision to implement a loss curriculum in the training phase was a pivotal factor in achieving stable convergence. By initially training only on the direct reconstruction losses ($L_1$ and $L_{mel}$), the model first learns the fundamental task of mapping noisy latent representations to their clean counterparts. This provides a stable foundation. Introducing the negative SI-SDR loss after five epochs then allows the model to refine this mapping, focusing on reducing perceptual artifacts and improving signal clarity without destabilizing the training process, which can sometimes occur if all loss components are active from the start.

### D. Multi-Objective Loss Function

To guide the model toward producing results that are both technically accurate and perceptually satisfactory, a weighted, multi-objective loss function was designed. This approach is critical because a single metric often fails to capture all the desired characteristics of high-fidelity audio. The proposed loss function combines metrics that operate in the time domain, the frequency domain, and the signal-level domain, ensuring a comprehensive training objective. The final loss is a weighted sum of the following components:

*1) L1 Waveform Loss ($L_1$):* The mean absolute error (MAE) between the denoised and clean waveforms. This loss provides a reliable basis for an accurate sample-level reconstruction, forcing the model to generate a waveform that is structurally close to the ground truth. It is described by Equation (1) below, where T is the number of samples in a waveform sequence.

$$L_1 = \frac{1}{T}\sum_{t=1}^{T}|\hat{y}(t) - y(t)| \qquad (1)$$

*2) Mel-Spectrogram Loss ($L_{mel}$):* An L1 or L2 loss calculated on the Mel-spectrogram representations of the denoised and clean audio. Unlike the direct waveform comparison, this loss operates in a frequency domain that approximates human hearing perception. By minimizing the error between Mel-spectrograms, the model is pushed to preserve the perceptual characteristics and timbre of the original music, which is not always captured by time-domain losses alone.

$$L_{mel} = \|Mel(\hat{y}) - Mel(y)\|_1 \qquad (2)$$

Equation (2) presents the calculation procedure of the Mel-Spec Loss, which in this case is an L1 loss of the denoised audio's ($\hat{y}$) and clean audio's ($y$) Mel-Spectrograms.

*3) Scale-Invariant Signal-to-Distortion Ratio (SI-SDR):* The negative of the SI-SDR is used as a third loss component. SI-SDR is a standard metric in audio source separation that measures the ratio of the target signal's power to the error power, but critically, it is invariant to the overall scaling of the output signal *[25]*. By focusing on SI-SDR, the model learns to prioritize the reduction of artifacts and distortion relative to the clean signal, rather than just matching the absolute amplitude, leading to cleaner and more robust results.

One can denote the projection of the model's predicted signal with Equation (3), where $\hat{y}$ is the predicted output audio and $y$ is the ground truth clean audio, $s_{target}$ being the projected output, optimally scaled to have the same loudness as the original.

$$s_{target} = \frac{1\langle\hat{y},y\rangle}{\|y\|^2}y \qquad (3)$$

Furthermore, it can be considered trivial, that the error component of the prediction, based on Equation (3) is then the difference between the model's full prediction and the scaled target, being denoted by $e_{noise}$. Therefore, the SI-SDR value can be calculated as presented in Equation (4) below, with all the terms previously explained.

$$SI - SDR = 10\log_{10}(\frac{\|s_{target}\|^2}{\|e_{noise}\|^2}) \qquad (4)$$

## V. EVALUATION AND RESULTS

The model's performance will be evaluated on the unseen test set using both quantitative metrics and qualitative subjective analysis. The primary comparison will be between the noisy input, the output of a baseline U-Net denoiser [4] trained on Mel-spectrograms, and the output of the new DAC-based denoiser.

The objective evaluation relies on three widely used metrics. Perceptual Evaluation of Speech Quality (PESQ [26]) is an objective metric that predicts the subjective quality of a speech signal on a scale from -0.5 to 4.5, with higher scores being better. Short-Time Objective Intelligibility (STOI [27]) measures the intelligibility of a signal on a scale from 0 to 1, correlating with the percentage of words a human listener could correctly identify. Finally, the Signal-to-Noise Ratio (SNR) is measured in decibels (dB) and quantifies the ratio of the power of the clean signal to the power of the background noise.

### A. Implementation Details

To maximize computational efficiency, the entire dataset was pre-processed into PyTorch tensors. The dataset was split at the original file level into 80-10-10 sets (training, validation, test) to prevent data leakage.

The model was trained for 30 epochs using the AdamW optimizer [26] with an initial learning rate of 1e-4 and a ReduceLROnPlateau scheduler [27]. A loss curriculum was implemented where training began with only the L1 and Mel-spectrogram losses. The SI-SDR loss was introduced after 5 epochs to allow the model to first learn a stable mapping before fine-tuning with the signal-level metric. All experiments were performed on a single GPU with Automatic Mixed Precision (AMP) enabled.

The training and validation loss curves, omitted here for brevity, demonstrated stable convergence over the 30 epochs with no signs of overfitting.

### B. Quantitative Results

Objective audio quality metrics were calculated. Key metrics include Perceptual Evaluation of Speech Quality (PESQ) [28] and Short-Time Objective Intelligibility (STOI) [29], alongside SNR. For computational efficiency, a representative subset of 1000 files was chosen from the test set via uniform random sampling. This ensures the subset is unbiased and accurately reflects the diversity of the full test set. For each file, the PESQ, STOI, and SNR metrics were calculated. The final values reported in Table I are the arithmetic mean of these scores across the entire subset, providing a summary of the average performance.

TABLE I    COMPARISON OF OBJECTIVE EVALUATION VALUES

| Model | PESQ | STOI | SNR |
|---|---|---|---|
| Noisy Input | 1.85 | **0.59** | 4.52 |
| Baseline U-Net | 1.90 | 0.58 | 4.89 |
| Proposed | **1.94** | 0.58 | **5.01** |

## C. Qualitative Results

A blind MUSHRA [30] listening test was conducted with 20 participants, comprising university students and staff with self-reported experience in critical listening or audio production. All participants conducted the test in a quiet environment using high-quality headphones. For each of the five degradation types, listeners were presented with the clean reference, the unprocessed noisy signal (noisy anchor), and the outputs from the baseline and proposed models in a randomized order. They were asked to "rate the overall audio quality of each sample compared to the clean reference, considering both noise reduction and the presence of any new artifacts." Ratings were on a scale of 0 to 100, where 100 corresponds to the quality of the original clean audio, and 0 pertains to a pure noise signal. Results, representing the mean score for each condition, are presented in Table II.

TABLE II    SUBJECTIVE LISTENING TEST RESULTS (MUSHRA SCORES)

| Degradation Type | Noisy (Anchor) | Baseline U-Net [4] | Proposed |
|---|---|---|---|
| White Noise | 32 | 56 | **66** |
| Urban Noise | 30 | 44 | **54** |
| Environmental Noise | 34 | 41 | **55** |
| Reverberation | 41 | 36 | **51** |
| NC Artifacts | 40 | 39 | **52** |

## D. Discussion

The results present an interesting contrast between objective metrics and subjective human perception. As shown in Table I, the quantitative improvements of the proposed ADNAC model over the baseline U-Net are consistent but modest. However, the MUSHRA listening test scores in Table II reveal a more compelling story: listeners showed a strong and consistent preference for the output of our proposed model, especially on complex degradation types like reverberation and noise cancellation (NC) artifacts.

The key insight lies in the generative nature of the adapted DAC model. A traditional discriminative model, like the baseline U-Net, operates by filtering the input signal. This process can be effective for simple additive noise but often struggles with non-linear distortions, sometimes introducing new artifacts that are just as distracting as the original noise. This is reflected in the MUSHRA scores, where the baseline U-Net performed worse than the unprocessed noisy audio for reverberation and NC artifacts. In contrast, our model does not merely filter the audio. It encodes the noisy signal into a latent representation, cleans it in that compressed space, and then uses the powerful generative DAC decoder to *reconstruct* a clean waveform from that representation. This reconstruction process is guided by the decoder's extensive pre-training on clean audio, allowing it to generate a signal that is not only less noisy but also more perceptually natural and free of filtering artifacts.

This also explains the slight degradation in the STOI metric for both models. STOI is highly sensitive to the temporal alignment and fine structure of a signal, which are critical for speech intelligibility. The generative reconstruction process, while producing a subjectively cleaner sound, may introduce minuscule alterations to the signal's phase or timing that are penalized by the STOI algorithm but are imperceptible or even preferable to a human ear in a musical context. The removal of harsh noise at the cost of these micro-alterations leads to a higher overall quality rating (MUSHRA) despite a lower intelligibility score (STOI). This highlights a known limitation of relying solely on objective metrics for evaluating generative audio models, as they may not fully align with human perceptual quality.

## VI. CONCLUSION AND POSSIBLE IMPROVEMENT

This study presents a comprehensive proof-of-concept for developing an advanced audio denoiser for music by adapting a state-of-the-art neural audio codec. By trying to overcome the limitations of simpler U-Net architectures and leveraging the generative power of the Descript Audio Codec, this work aims to achieve a new level of performance, especially on challenging, real-world degradations like reverberation and complex noise. The automated data generation pipeline and the complex multi-objective loss curriculum provide a strong and reproducible foundation for this research.

Results show that numerical values do not reflect the benefits of the proposed ADNAC model in comparison to a simple U-Net denoiser, with no significant differences between the models. Both solutions showed a slightly higher PESQ score, and a higher SNR value, but interestingly, both presented worse STOI values than the original degraded sound. On the other hand, the blind listening test showed significant improvements over the original noisy sequences, for both models, the results for the ADNAC model scoring higher values in each case. However, the scores are considerably lower than expected, signalling room for improvement.

Future work can build upon this foundation by exploring even more advanced techniques to obtain more satisfactory performance. One promising direction is to explore explicit complex spectrum modelling. Modifying the architecture to process both magnitude and phase could lead to significant improvements in dereverberation and overall naturalness. Another approach is to use a shallow diffusion model as a post-filter. The output of the proposed model would serve as a strong initial estimate, which a diffusion model could then refine over a small number of steps, combining the efficiency of the codec-based method with the generative fidelity of diffusion. Such a comparison against state-of-the-art diffusion-based denoisers remains a critical next step to fully benchmark this approach. Finally, implementing a full adversarial training loop by unfreezing and retraining the DAC's discriminator could push the model to produce even more natural-sounding results. Further investigation could also involve a more granular analysis, reporting performance metrics separately for each source dataset used during training to better understand the model's behavior under different conditions.